# Coupling between phonons and magnetic excitations in orthorhombic $Eu_{1-x}Y_xMnO_3$

J. AGOSTINHO MOREIRA, A. ALMEIDA, W. S. FERREIRA AND M. R. CHAVES

*Departamento de Física, IFIMUP and IN-Institute of Nanoscience and Nanotechnology, Faculdade de Ciências, Universidade do Porto, Rua do Campo Alegre, 687, 4169-007 Porto, Portugal.*

J. KREISEL

*Laboratoire des Matériaux et du Génie Physique. CNRS. Grenoble Institute of Technology, 38016. Grenoble. France.*

S. M. F. VILELA AND P. B. TAVARES

*Centro de Química. Universidade de Trás-os-Montes e Alto Douro. Apartado 1013, 5001-801. Vila Real. Portugal.*

* electronic address: jamoreir@fc.up.pt

## Abstract

In this work we present a detailed study of the structural and lattice dynamic properties of Y-doped $EuMnO_3$ ceramics ($Eu_{1-x}Y_xMnO_3$, with $0\leq x\leq 0.5$). A thorough analysis towards the correlation between both structural and Raman modes parameters has been undertaken. Our results provide evidence for two main structural distortions of $MnO_6$ octahedra, arising from a cooperative Jahn-Teller and rotational distortions in these compounds. The temperature dependence of the $B_{1g}$ symmetric stretching mode of the $MnO_6$ units has revealed either a positive or negative shift regarding the pure anharmonic temperature dependence of the phonon frequency, which strongly depends on the Y-concentration. This frequency renormalization is explained in terms of a competition between ferro and antiferromagnetic interactions. The pronounced frequency renormalization of the symmetric stretching mode for x=0.2 evidences the predominance of the ferromagnetic interactions against the antiferromagnetic ones, which is in good correlation with the spontaneous magnetization measured in this compound. The emerging of the shifts referred well above the Néel temperature is likely associated with the coupling between phonons and dynamical fluctuations of the magnetic system, which provides further grounds for the existence of strong spin-phonon coupling in this compound.

## I. INTRODUCTION

Magnetoelectric multiferroic materials, exhibiting coupled ferroelectric and ferromagnetic orders parameters, are attractive for future technological applications, such as a new generation of both electrically and magnetically controlled multifunctional devices. Unfortunately, the number of such materials remains scarce.[1] On the other hand, materials presenting coupled ferroelectric and antiferromagnetic properties are more frequent, and they are just known as magnetoelectric materials.[2] Among them, rare-earth manganites have drawn an important interest in the scientific community due to their rich phase diagram and the associated microscopic mechanisms of which the understanding remains challenging.

In conventional ferroelectrics, such as $BaTiO_3$, hybridization between the titanium 3d states and the oxygen 2p states is essential for proper ferroelectricity.[3] Ferroelectricity in magnetoelectric rare-earth manganites is entirely different in nature, as it has an improper character, which has been attributed to the spin-lattice interactions in a modulated off-centre symmetric magnetic structure.[4,5,6] In magnetoelectric rare-earth manganites an electric polarization has been only observed when they are cooled under high electric fields (E > 1 kV/cm) [7], and the saturation value of the polarization is rather small. It has been proposed that ferroelectricity can be originated from a variety of spiral magnetic structures[4], and can be explained in terms of the inverse Dzyaloshinski-Morya model, where the electric polarization is expected as[8.9]:

$$\vec{P} = \sum_{i,j} A\vec{e}_{ij} \times (\vec{S}_i \times \vec{S}_j) \cdot \qquad (1)$$

$\vec{e}_{ij}$ denotes the unit vector connecting the interacting neighbour $\vec{S}_i$ and $\vec{S}_j$ spins, and A is the coupling constant between electric polarization and magnetic momenta. As the electric polarization arises from lattice distortions, the study of the spin-phonon coupling is particularly notorious in the systems that present simultaneously magnetic and ferroelectric properties. However, spin-phonon, and thus magnetoelectric coupling has shown to be rather weak in to date studied materials, inhibiting their use in technical applications. Consequently, from both fundamental and application point of view, a further understanding of spin-phonon coupling remains an important issue.

A Raman scattering study of orthorhombic rare-earth manganites ($ReMnO_3$) revealed a significant coupling between spins and lattice for compounds with Re = Pr, Nd and Sm near and below the Néel temperature.[10] We note that the effect of the magnetic ordering is very weak or negligible in rare-earth manganites with Re = Gd, Tb, Dy, Ho and Y, which are considered as magnetoelectric compounds.[10] Any comparative analysis of these compounds is rather complex, because both the variations of the rare-earth radii and the different values of magnetic moment in different Re ions have to be taken into account.[11] This complexity is reduced in the $Eu_{1-x}Y_xMnO_3$ system ($0 \leq x \leq 0.5$), with an orthorhombic distorted perovskite structure of P*bnm* symmetry, because the magnetic properties are entirely due to the manganese 3d spins, since $Eu^{3+}$ and $Y^{3+}$ present no magnetic momenta.[12] Further interest in this system comes from its $(x,T)$ phase diagram, which shows a rich phase sequence at low temperatures, exhibiting different types of both polar and magnetic structures.[13] Due to the fact that the the $Y^{3+}$ radius (1.02 Å) is smaller than that of $Eu^{3+}$ (1.09 Å), the orthorhombic distortion of the $Eu_{1-x}Y_xMnO_3$ increases with x, and can be quantified by the spontaneous orthorhombic strain factor:[14]

$$e = \frac{2(b-a)}{b+a} , \qquad (2)$$

where $a$ and $b$ denote the values of the lattice parameters. This distortion is a consequence of the Jahn-Teller cooperative effect and the tilting of the octahedra around the $a$-axis, thus lowering the symmetry of the system.[14] As a consequence of this increasing lattice deformation, the orbital overlap becomes larger via the Mn-O(I)-Mn bond angle and, in turn the electronic properties of the system are modified.[15]

In this paper we present a detailed study of the crystal structure and lattice dynamics at room temperature, and of the spin-phonon coupling in $Eu_{1-x}Y_xMnO_3$ in the concentration range $0 \leq x \leq 0.5$, by using both X-ray powder diffraction and Raman spectroscopy. In particular, we aim at correlating the Raman-active modes with the lattice distortions in this system, and to determine the relevance of the interaction between magnetic spins and internal vibrations to the onset of electric polarization.

## II. PHASE DIAGRAM OF $Eu_{1-x}Y_xMnO_3$

Let us first discuss the literature work which is relevant for the interpretation and understanding of our own experimental results.

Based on magnetic and dielectric measurements, as well as on heat capacity studies a detailed ($x,T$) phase diagram has been proposed for $Eu_{1-x}Y_xMnO_3$, which is shown in Fig. 1.[13] The paramagnetic phase above the Néel temperature, $T_N \approx 50$ - 45 K, is followed by an antiferromagnetic phase (AFM1). By analogy with others $ReMnO_3$ compounds (Re = Ho, Dy, Tb, Gd, Eu), Hemberger *et al*[13] proposed a sinusoidal modulation for the AFM1 phase, which was confirmed experimentally by Yamasaki *et al*[16], for $x$ = 0.2, 0.3 and 0.4, using synchrotron radiation. For $x < 0.15$, a new antiferromagnetic phase, with weakly ferromagnetic character, is established below $T_1 \approx$ 42 - 30 K.[13,16] Taking into account the absence of ferroelectric polarization in this magnetic phase, it has been proposed that an A-type collinear spin structure is established.[13,16] For $0.15 \leq x < 0.3$, the system exhibits simultaneously a canted antiferromagnetism and an electric polarization along the *a*-axis below $T_1 \approx 30$ K.[13] Yamasaki *et al*[16] showed experimentally, that for the special case of $x$ = 0.2, the superlattice reflections observed in the AFM1 phase disappear below 30K. Based on this result, these authors have proposed a ferromagnetic alignment of the Mn spins in the *ab*-plane, without modulation, coupled antiferromagnetically along the *c*-axis, where the magnetization appears.[16] Contradictory results have been obtained for the polar character of the low temperature magnetic phase for $x$ = 0.2. While Hemberger *et al*[13] found a finite electric polarization, Yamasaki *et al*[16] reported the absence of any polar order in samples when measured with silver paste electrodes and under an electric field of 2 kV/cm. Yamasaki *et al*[16] assigned the discrepancy between their result and the one reported by Hemberger *et al*,[13] to the non-stoichiometry of the samples studied by these authors. For $x > 0.25$, a modulated antiferromagnetic phase (AFM2) is observed below $T_1 \approx 28$ - 22K, along with a ferroelectric polarization. The modulation wave-vector depends on the Y-content. According to the results obtained by Yamasaki *et al*[16], for $x$ = 0.3, a commensurate antiferromagnetic phase ($q_l$ = 0.5) is observed, while for $x$ = 0.4 an incommensurate antiferromagnetic phase, with $q_l \approx 0.56$ is established. For $0.4 \leq x \leq 0.5$, an electric polarization along the *a*- or the *c*-axis, depending on the temperature of the sample, is observed. The ferroelectricity in $Eu_{1-x}Y_xMnO_3$ ($x \geq 0.2$)may have an improper character, and could be attributed to a spin-lattice interaction in the modulated off-centre symmetric magnetic structure.

The analysis of the experimental study of infrared transmittance in the terahertz and far infrared regions (3 $cm^{-1} \leq \omega \leq 250$ $cm^{-1}$) of $Eu_{1-x}Y_xMnO_3$ (with $x$ = 0.2, 0.3 and 0.5) evidenced clear anomalies in the dielectric constant around 35 K.[17,18] The magnitude of this effect is associated with the strength of the coupling between optically-active phonons and spin waves. These excitations interact strongly with electromagnetic waves of adequate wavelengths and are called electromagnons. Electromagnetic waves can provide in $Eu_{1-x}Y_xMnO_3$ a contribution to the static dielectric constant in magnetic ordered phases. For $x$=0.25 the electromagnons observed are suppressed by external magnetic fields, which induce a canted antiferromagnetic phase.[17] In addition to electromagnons a broad contribution in the infrared transmission spectra is observed for $x = 0.25$ in the definite polarization e||*a* and h||*b*, and persists to temperatures as high as 150 K.[17] Above $T_1$, the low-frequency response has origin in this background. At T=10 K, the background accounts for approximately 50% for the total low-frequency oscillator strength below 140 $cm^{-1}$.[17] The gradual decrease of this background with temperature above $T_N$ suggests that magnetic fluctuations are the origin of this behaviour. It has been proposed that this background arises from the coupling of the phonons to dynamic fluctuations of the magnetic system in the paramagnetic phase.[17]

## III. EXPERIMENTAL DETAILS

Details of the sample processing are available in Ref. 19. The phase purity and the crystallographic characterization of the ceramic samples were checked using X-ray powder diffraction and scanning electron microscopy with energy dispersive spectroscopy. Scanning electron microscopy analysis reveals in both systems a typical ceramic microstructure with regular shaped crystal grains ranging from 3 up to 10 μm in diameter. The Rietveld method was used to analyse the room temperature X-ray spectra. This analysis has enabled us to calculate the cell parameters, cell volume, bond lengths and bond angles as a function of the Y-concentration.

The Raman scattering studies were performed with polished pellets ($3\times4\times5$ $cm^3$). The samples were placed in a closed-cycle helium cryostat (10-300 K temperature range) with a temperature stability of about $\pm0.2$ K. The temperature homogeneity in the samples was achieved with a cooper mask set-up. The temperature of the sample was estimated to differ by less than 1 K from the temperature measured with a silicon diode attached to the sample holder.

The unpolarized Raman spectra of $Eu_{1-x}Y_xMnO_3$ have been measured in the pseudo-backscattering geometry, on cooling runs. The 632.8 nm polarized line of a He-Ne laser was used for excitation, with an incident power of about 5 mW impinging on the sample. The scattered light was analyzed using a T64000 Jobin-Yvon spectrometer, operating in triple subtractive mode, and equipped with liquid nitrogen cooled CCD and photon-counting device. Identical conditions were maintained for all scattering measurements. The spectral slit width was about 1.5 cm$^{-1}$.

The sum of independent damped harmonic oscillators, according to the general formula:[20]

$$I(\omega,T) = (1 + n(\omega,T))\sum_{j=1}^{N} A_{oj} \frac{\omega \Omega_{oj}^2 \Gamma_{oj}}{(\Omega_{oj}^2 - \omega^2)^2 + \omega^2 \Gamma_{oj}^2}. \quad (3)$$

was fitted to the experimental data. Here $n(\omega,T)$ is the Bose-Einstein factor; $A_{oj}$, $\Omega_{oj}$ and $\Gamma_{oj}$ are the strength, wavenumber and damping coefficient of the j-th oscillator, respectively.

## IV. EXPERIMENTAL RESULTS

### a. Structure and lattice dynamics at room temperature.

#### a.1. X-ray diffraction studies.

Figure 2 shows an ORTEP plot of the unit cell of $EuMnO_3$ at room temperature. The $Eu_{1-x}Y_xMnO_3$ crystals, with $x \leq 0.5$, present the same *Pbnm* orthorhombic structure, with Z = 4, at 300 K, and are formed by a network of corner-sharing $MnO_6$ octahedra developing chains along the *c*-axis, belonging to the family of rotationally distorted perovskites. The $Eu^{3+}$ or $Y^{3+}$ ions occupy the interstices between octahedra.

The cell parameters and the cell volume as a function of *x*, obtained from the analysis of the x-ray diffraction data, are shown in Figure 3(a). For all samples here studied, the lattice parameters fulfil the $c/\sqrt{2} < a < b$ relation, which is characteristic of the so-called O' structure, typically found in other rare-earth manganites presenting distortions of the octahedral environment of the $Mn^{3+}$ ions, associated with a strong Jahn-Teller distortion of the $MnO_6$ units and orbital ordering.[14,21] The increase of Y-concentration with a smaller ionic radius than $Eu^{3+}$ leads to a decrease of both cell parameters and volume, actually due to the decrease of the effective A-site volume. As it can be seen from Figure 3(a), the slope of the straight line *b*(x) parameter ($\Delta b/\Delta x$ = -0.04±0.01Å) is smaller than the ones of *a*(x) and *c*(x) ($\Delta a/\Delta x$ = -0.084±0.001 Å, $\Delta c/\Delta x$ = -0.067±0.003 Å), as it has been observed in other rare-earth manganites. This kind of behaviour has been attributed to the tilting of $MnO_6$ octahedra in *Pbnm* perovskite structure, of the type $a^-a^-c+$ in Glazer´s notation,[22] for which the distortion driven by a reduction of the A-site volume leaves *b* slightly *x*-dependent. The spontaneous orthorhombic strain (*e*) as a function of *x* is depicted in Figure 3(b). The *e* parameter, which characterizes the orthorhombic distortion of the lattice (Eq. 2), increases monotonically with *x*, corroborating the increasing deformation of the lattice with respect to the ideal cubic perovskite structure. The increase of the spontaneous orthorhombic strain *e* provides a clear evidence for a continuous octahedra tilting and also measures the distortion of the octahedra: both effects are simultaneously observed in several $ReMnO_3$, which lead to changes of the lattice parameters, and significant distortions of the Mn-O(I)-Mn angles.[23] The observed decrease of the cell volume scales very well with the effective A-site volume, as it can be seen in Figure 3(c). It is interesting to remind that the decreasing of the cell volume strongly influences the magnetic exchange and the electronic orbital overlapping and, in this way, the magnetic properties of these materials, as it has been observed in other rare-earth manganites.

From Rietveld refinement of the atomic positions, we have calculated the values of both the length and angle of chemical bonds involved in the $MnO_6$ octahedra, the R-O(I) distances and the octahedra tilt angle for each value of *x*. Figure 4 shows the R-O(I) and Mn-O bond lengths, and the octahedra tilt angle as a function of the Y-concentration.

Two different R-O(I) lengths have been detected for all the considered compositions. Whilst for low Y-concentrations ($x \leq 10$) the R-O(I) lengths are practically constant, a slightly decrease is observed as x increases for higher Y-concentrations. In the range $0.15 \leq x \leq 0.4$, the difference between the two R-O(I) length values is more pronounced. The difference between both R-O(I) bond length values is likely associated with the behaviour of the octahedra tilt angle.

Significant distortions in the $MnO_6$ unit are detected from the analysis of the values of the Mn-O bond lengths. In fact, from the Rietveld refinement, we have found three different values for the Mn-O bond length for each composition, evidencing the existence of deformations of the $MnO_6$ unit. While only one value is found for the Mn-O(I) bond lengths, two different values for the Mn-O(II) ones are displayed, whose difference is about 20%. This result is also observed in other rare-earth manganites and is a consequence of the Jahn-Teller distortions which manifest themselves by a rather pronounced difference between the values of the Mn-O(II) bond lengths, roughly lying in the ab-plane. In order to quantify the distortion of the octahedra, we have used the $\Delta_d$ parameter, defined as:[14]

$$\Delta_d = \frac{1}{6}\sum_{i=1}^{6}\left[\frac{d_i - <d>}{<d>}\right]^2 . \qquad (4)$$

where $d$ = Mn-O and $<d>$ is the mean value of Mn-O. Using the data obtained in this work, we have determined $\Delta_d = 67.3\times10^{-4}$, which is a significantly larger than the values obtained for other orthorhombic $ReMnO_3$ compounds (see Table I).[24]

The tilt, defined as [180º-(Mn-O(I)-Mn)]/2,[25] reveals itself to be an important parameter as its value reflects the magnetic interactions responsible for the spin structure exhibited by the rare-earth manganites. As we can see from Figure 4(c), the tilt angle increases monotonously with increasing $x$. The increase of the tilt angle with decreasing cell volume corroborates the continuous deviation from the cubic structure due to Y-doping obtained in the previous results. It is interesting to stress that the values of the tilt angle observed in both $TbMnO_3$ and $DyMnO_3$, which exhibit strong magnetoelectric effects, are comparable with those obtained for $x$ ~ 0.15 and 0.2, respectively.[14]

a.2. Raman spectroscopic studies.

Unpolarized Raman spectra of $Eu_{1-x}Y_xMnO_3$, $x$ = 0, 0.1, 0.2, 0.3, 0.4 and 0.5, recorded at room temperature are shown in Figure 5.

In the orthorhombic rare-earth manganites, the activation of the Raman modes is due to deviations from the ideal cubic perovskite structure. Factor group analysis of the $EuMnO_3$ structure provides the following decomposition corresponding to the 60 normal vibrations at the Γ-point of the Brillouin zone:

$$\Gamma_{acustic} = B_{1u}+B_{2u}+B_{3u}$$

$$\Gamma_{optical} = (7A_g+7B_{1g}+5B_{2g}+5B_{3g})_{Raman\text{-}active} + (8A_u+10B_{1u}+8B_{2u}+10B_{3u})_{IR\text{-}active}.$$

The Raman-active modes preserve the inversion centre, thus $Mn^{3+}$ ions do not contribute to the Raman spectra. Due to the polycrystalline nature of the studied samples, our Raman spectra exhibit simultaneously all Raman-active modes $A_g$, $B_{1g}$, $B_{2g}$ and $B_{3g}$. Earlier reports by Lavèrdiere *et al*,[10] propose that the more intense Raman bands are of $A_g$ and $B_{2g}$ symmetry. In good agreement with this, our observed $A_g$ and $B_{2g}$ modes are the more intense bands in $Eu_{1-x}Y_xMnO_3$. Contrarily to other ceramic systems, like $SrTiO_3$, the activation of infrared modes due to the symmetry breaking in the grain boundaries is not observed in our Raman spectra.[26] Figure 5 shows that the spectral signature of all $Eu_{1-x}Y_xMnO_3$ (with $x \leq 0.5$) compounds is qualitatively similar in the 300-800 $cm^{-1}$ frequency range in terms of frequency, linewidth and intensity. The similarity of the Raman signatures suggests that all studied powders present at ambient conditions the same space group and that the internal modes of the $MnO_6$ octahedra units are not substantially affected by the Y-doping.

Nevertheless, a closer quantitative analysis of the spectra shows subtle changes as a function of the substitution rate. For instance, the broad feature located at around 520 $cm^{-1}$ becomes more pronounced as the Y-concentration increases. Also, the frequency of the band located close to 364 $cm^{-1}$ does significantly increase with increasing *x*.

A detailed study, published by L. Martín-Carrón *et al*,[25] concerning the dependence of the frequency of the Raman bands in some stoichiometric rare earth manganites $ReMnO_3$, enables us to assign the more intense Raman bands in our spectra. In $Eu_{1-x}Y_xMnO_3$ the band at 613 $cm^{-1}$ is associated with a symmetric stretching mode involving the O(II) atoms (symmetry $B_{2g}$) , [25,27,28,29] the band at 506 $cm^{-1}$ to a bending mode (symmetry $B_{2g}$), the band at 484 $cm^{-1}$ to a Jahn-Teller type asymmetric stretching mode involving also the O(II) atoms (symmetry $A_g$), and the band at 364 $cm^{-1}$ to a bending mode of the tilt of the $MnO_6$ octahedra (symmetry $A_g$).

On the basis of these assignments, it is now interesting to correlate the *x*-dependence of the frequency of these Raman bands with the structural changes induced by the Y-doping. The more noticeable stretching modes in $ReMnO_3$ are known to involve nearly pure Mn-O(II) bond and they are found to be slightly dependent on the chemical pressure. In orthorhombic rare-earth manganites, the stretching modes change less than 5 $cm^{-1}$ with the rare-earth substitution, from La to Dy.[27] Figure 6 shows the evolution of the two bands located around 613 $cm^{-1}$ and 484 $cm^{-1}$. The observed frequency changes of only 2 $cm^{-1}$ when *x* increases from 0 to 0.5, correlates well with a weak dependence of the Mn-O(II) bond lengths with x (see Figure 3(b)). The weak *x*-dependence of the frequency of these modes provides further evidence for a slight dependence of the $MnO_6$ octahedron volume and Mn-O bonds lengths on the Y-doping, in agreement with literature work on other rare-earth manganites.[25] The remaining two modes *B* and *T* depicted in Figure 6, exhibit a large variation with $x$ (about 10-15 $cm^{-1}$ when $x$ varies from 0 to 0.5), which correlates rather well with the x-dependence of the tilt angle. The lower frequency *T* mode exhibits the largest variations with $x$, and it is actually associated with an external mode $A_g$ with origin in the tilt mode of the $MnO_6$. A linear dependence of the frequency of *T* mode in the tilt angle is observed from Figure 7. The slope of the linear relation is 5 $cm^{-1}$/deg, which is a very small value when compared with the one obtained for the same mode in other orthorhombic manganites (23 $cm^{-1}$/deg).[27] Mode *B* is assigned to the bending mode $B_{2g}$ of the octahedra.[27] The two broad shoulders observed at round 470 $cm^{-1}$ and 520 $cm^{-1}$ are likely the $B_{2g}$ in-phase O(II) scissorlike and out-of-plane $MnO_6$ bending modes.

The increase of Y concentration provides evidence for some modifications in the crystal structure, which manifest themselves through changes of the mode frequencies, relative intensities, and linewidths. The most interesting example concerns the intensity interchange between the bending and asymmetric stretching modes (spectral range 470 - 520 $cm^{-1}$) and the damping coefficient of the symmetric stretching mode. The ratio between the intensity of the bending and the asymmetric stretching modes is presented in Figure 8, where a linear decrease of the intensity ratio as the Y-concentration increases is observed. This effect has been also observed in the same spectral range for the whole rare-earth manganite series, and it has been associated with the coupling between these two modes, which is stronger for Re = Eu, Gd and Tb.[27] The inset of Figure 8 shows the linewidth of the symmetric stretching modes as a function of the Y-concentration. An increase of the linewidth with increasing $x$ is observed due to the structural disorder arising from the partial substitution of $Eu^{3+}$ by $Y^{3+}$.

### b. Temperature dependence of the Raman spectra.

Figure 9 shows the unpolarized Raman spectra of $EuMnO_3$, $Eu_{0.8}Y_{0.2}MnO_3$ and $Eu_{0.5}Y_{0.5}MnO_3$, recorded at 200K and 9K. As it can be seen in Figures 9(a) to 9(c), the spectra at 200K and 9K show only very small changes in their profiles. Particularly, no new bands were detected at low temperatures. The absence of well-defined activated Raman bands, even for the composition where a spontaneous ferroelectric order is expected, may have origin in two different mechanisms: the maintenance of the inverse centre or the weak polar character of the ferroelectric phases for $x \geq 0.2$.

Nevertheless, the detailed analysis of the spectra reveals the presence of anomalies in the temperature dependence of some phonon parameters across the magnetic phase transitions. As an example, Figure 10 shows the Raman spectra of $Eu_{0.8}Y_{0.2}MnO_3$, recorded at several fixed temperatures, in the 590-640 $cm^{-1}$ spectral range. As we can see from Fig. 10, the frequency of the symmetric stretching mode exhibit intriguing temperature behaviour, hardening as the temperature decreases from room temperature towards 100K, then softening as the temperatures further decreases. Such a behaviour provides evidence for some kind of structural rearrangement or coupling.

According to the spin-phonon coupling models, one should expect detectable changes in the phonon frequencies on entering the magnetic phases, reflecting the phonon renormalization, proportional to the spin-spin correlation function for the nearest $Mn^{3+}$ spins. Aiming at searching for a spin-phonon coupling in $Eu_{1-x}Y_xMnO_3$ series, we have monitored in detail the temperature dependence of the parameters characterizing the $MnO_6$ Raman-active modes. Among them, we have found that the symmetric stretching mode (SS) around 615 $cm^{-1}$ is the most sensitive to the magnetic order. In fact, this mode affects the geometrical parameters associated with the spontaneous orthorhombic strain *e* and, so a strong coupling between the SS mode and the electronic degrees of freedom is expected. In other rare-earth manganites, like Ca-doped $PrMnO_3$, this mode is so strongly coupled with the electronic system that it can be used to control a metal-insulator transition, by its coherent manipulation through selective mode excitation.[30]

The temperature dependence of the frequency of the symmetric stretching mode for the different studied chemical compositions is shown in Figure 11, together with the insets, which present the temperature dependence of the corresponding linewidths.

In order to calculate the influence of the magnetic exchange interactions on the phononic behaviour, we describe the purely anharmonic temperature dependence of the frequency and of the linewidth of the different modes by the model:[31]

$$\omega(T) = \omega(0) + C\left(1 - \frac{2}{e^x - 1}\right), \qquad (5)$$

for the temperature dependence of the frequency of the transverse mode and:

$$\Gamma(T) = \Gamma(0)\left(1 + \frac{2}{e^x - 1}\right) \tag{6}$$

for the temperature dependence of its linewidth. In equations (5) and (6), ω(T) and ω(0) is the frequency of the optical mode at the temperature T and 0K, respectively, $x = \frac{\hbar\omega_o}{2k_B T}$ where $\omega_o$ is the characteristic frequency of the mode, Γ(T) and Γ(0) are the linewidths of the mode at the temperature T and 0K, respectively. The solid lines in Figure 11 correspond to the best fit of these equations to the high-temperature range data (T > 100K), with the adjustable parameters C, ω(0),Γ(0) and $\omega_o$.

The results displayed in Figure 11 clearly show that for $Eu_{1-x}Y_xMnO_3$, with $x$ = 0, 0.3 and 0.4, there is only a faint frequency shift at $T_N$ in the temperature dependence of the phonon frequency from the normal anharmonic behaviour. In contrast to this, a significant negative frequency shift is found for $x$ = 0.2 and an observable negative and positive shifts are observed for $x$ = 0.1 and $x$ = 0.5, respectively.

For all these compounds the shifts appear well above the onset of the magnetic order and consequently it is very unlikely that these effects are driven by any kind of long range spin ordering. It may be that the shifts occur when the temperature allows some kind of local order concerning the spins which can be probed by Raman scattering, as it will be discussed at the end of this work.

The frequency shift of a given phonon as a function of temperature, due to the spin-phonon coupling, is determined by the spin-spin correlation function:[32,33]

$$\omega = \omega_o + \gamma\langle \vec{S}_i \cdot \vec{S}_j \rangle, \tag{7}$$

where ω is the renormalized phonon frequency at a fixed temperature, $\omega_o$ denotes the frequency in the absence of spin-phonon coupling, and γ is the spin-phonon coupling constant.

When there are ferromagnetic and antiferromagnetic competitive interactions it was proposed for the frequency shift:[33,34]

$$\omega - \omega_o \propto -R_1\langle \vec{S}_i | \vec{S}_j \rangle + R_2\langle \vec{S}_i | \vec{S}_k \rangle \tag{8}$$

where $R_1$ and $R_2$ are spin dependent force constants of the lattice vibrations. $R_1$ is associated with the ferromagnetic nearest neighbour and $R_2$ is associated with the antiferromagnetic next-nearest neighbour exchange. This model predicts negative or positive frequency shifts depending on the relative strength between the ferromagnetic and antiferromagnetic interactions.

In the grounds of the model presented to above, the experimental results displayed in Figure 11 can be tentatively interpreted by assuming the coexistence of ferromagnetic and antiferromagnetic competitive interactions in the $Eu_{1-x}Y_xMnO_3$ system, evidenced through the existence of modulated magnetic structures in the low temperature range, i.e., below 50 K.[16]

Regarding the phase diagram (*x*,*T*) of $Eu_{1-x}Y_xMnO_3$ system[13], we note that for $x = 0$, 0.1 and 0.2 the system presents a weak ferromagnetic phase at low temperatures. We have observed a non-linear M(H) relationship and a remanent magnetization for the $x = 0.2$ compound below T = 30 K, seen in Figure 12, evidencing indeed a weak ferromagnetic character of the $Eu_{0.8}Y_{0.2}MnO_3$. On the other hand, for $x = 0.4$ and 0.5 an antiferromagnetic character is found at low temperatures.[13,16] Consequently, the negative shift observed for $x$=0.1 and 0.2 is in close agreement with the ferromagnetic behaviour of these compounds, while the positive shift found for $x = 0.5$ shows that the antiferromagnetic interactions prevail for this *x*-content. We speculate that the absence of significant anomalies for $x = 0.3$ and $x = 0.4$ can be explained by the fact that the phonon signatures of antiferromagnetic and ferromagnetic interactions nearly compensate, and thus lead to an apparent reduced spin-phonon coupling.

Except for $EuMnO_3$, the linewidth deviates, around 100 K, from the purely anharmonic temperature dependence behaviour. For $x = 0.20$, the linewidth presents a further anomaly at $T_N \approx 50$K where the temperature derivative of the wave number is maximum.

A similar analysis of the other Raman bands was performed and the results show that the shifts are less pronounced for these modes than for the symmetric one.

Through out of the scope of this work, it is worth to note the kink at ~250 K in the temperature behaviour of the SS mode frequency for $x = 0.4$ (see Figure 11(e)). However, a detailed analysis of this issue will be presented in a forthcoming work.

## V. DISCUSSION AND CONCLUDING REMARKS

In this work we have presented a systematic study concerning the structure at room temperature and the phonon behaviour at low temperatures in orthorhombic $Eu_{1-x}Y_xMnO_3$ manganites.

The structural data obtained from both x-ray diffraction and Raman scattering experiments reveal orthorhombic deformations due to a strong cooperative Jahn-Teller distortion, which increases with increasing $Y^{3+}$ substitution rate. In the Raman scattering data, the lattice distortions manifest themselves by frequency shifts of both the tilt and the bending modes of the $MnO_6$ units, modifications of the mode coupling, enhancement of the energy transfer mechanisms between modes, and increase of the Raman band linewidths. The data analysis points at a correlation between the x-dependence of the more intense Raman bands with the Mn-O bond lengths, and the Mn-O(I)-Mn bond angle; i.e., the stretching modes (symmetric and antisymmetric) correlate to Mn-O bonds, while bending and tilt mode are associated with the Mn-O(I)-Mn angle.

The study of Raman spectra as a function of the temperature, in the mixed polycrystalline $Eu_{1-x}Y_xMnO_3$ ($0 \leq x \leq 0.50$) reveals a shift in the temperature dependence of the phonon frequency away from the normal anharmonic behaviour, particularly significant for $x = 0.20$. The frequency shift, due to the spin-phonon coupling interaction, regarding to the pure anharmonic temperature dependence of the phonon frequency, is positive or negative depending on the x values. This behaviour can be understood in terms of a competition between ferro- and antiferromagnetic interactions revealed through the existence of modulated structures in the low temperature phases of these materials. The marked renormalization of the frequency of the symmetric stretching mode involving the $MnO_6$ octahedra, for $x = 0.20$, suggests a predominance of ferromagnetic interactions in this compound, in agreement with the non-linear behaviour of the magnetization versus magnetic field. According to Hemberger *et al*,[13] the samples with this Y-content exhibit simultaneously a spontaneous magnetic momentum and ferroelectricity, which would yield a multiferroic character to the compound with $x = 0.2$.

As we will see in the following, our Raman results correlate very well with experiments of the terahertz transmittance on $Eu_{1-x}Y_xMnO_3$.[17,18] The analysis obtained from these experiments revealed the existence of electromagnons in samples with $x = 0.20$, 0.30 and 0.50, providing information about the static dielectric constant below $T_1$, and also providing evidenced for a broad background absorption observed up to $T_N$+50K.

In our Raman work, we have observed a deviation of the frequency from the anharmonic behaviour, well above the magnetic phase transition, which is at first sigth unexpected. Laverdière *et al*[10] also have observed a softening for $A_g$ and $B_{2g}$ stretching modes in, e.g. $NdMnO_3$ and $DyMnO_3$, starting well above $T_N$, which they relate to a small expansion in the Mn-O distance rather than to a spin-phonon coupling. However, if we consider that manganites are well-known for their often peculiar local structure in terms of electronic and magnetic properties, it is also plausible to associate this type of behaviour with some local structural disorder arising from local magnetic fluctuations, similarly to that was proposed in rare-earth nickelates, which also present local fluctuations.[36,37] As above $T_N$ no long-lived magnons are possible, Aguilar *et al*[17] have associated the existence of the background absorption in the terahertz region of the paramagnetic phase with some kind of coupling between phonons and dynamical fluctuations of the magnetic system. We have reconsidered the correlation between the total spectral weight below 140 $cm^{-1}$, depicted in Fig 2 of Ref. 17, and the frequency shift calculated from our results for $x$ = 0.20. The results are displayed in Figure 13. They provide support for a mechanism that involves spin-spin coupling, since the shift of the measured frequency (proportional to $\vec{S}_i \cdot \vec{S}_j$) for $x$ = 0.20 scales rather well with the spectral weight displayed in Figure 2 of the Ref. 17, in the temperature range 50K-100K. This spectral weight, calculated from the sample with $x$ = 0.25, is reported to be proportional to the coupling of phonons to the dynamical fluctuations of the magnetic system, in the paramagnetic phase, i.e. also proportional to $\vec{S}_i \cdot \vec{S}_j$. As expected, the scaling found is independent of yttrium content, but the factor of proportionality between the shift of frequency and the spectral weight is actually not.

**ACKNOWLEDGMENTS**

This work was supported by Fundação para a Ciência e Tecnologia, through the Project PTDC/CTM/67575/2006 and by Program Alβan, the European Union Program of High Level Scholarships for Latin America (scholarship no. E06D100894BR). J. Kreisel thanks the European STREP MaCoMuFi for financial support.

Figure Captions:

Figure 1. ($x$,$T$) phase diagram of $Eu_{1-x}Y_xMnO_3$ ($0 \leq x \leq 0.55$)obtained from the results described in Ref. 13.

Figure 2. ORTEP plot of the $EuMnO_3$. The A denotes the rare-earth or $Y^{3+}$ ions, the B denotes the $Mn^{3+}$ ions, O(I) and O(II) denote the apical and equatorial $O^{2-}$ ions, respectively.

Figure 3. (a) Lattice parameters and cell volume of $Eu_{1-x}Y_xMnO_3$, as a function of x. (b) Spontaneous orthorhombic strain as a function of x. (c) Cell volume versus the effective volume of the A-site.

Figure 4. Dependence of the R-O(I) (a), Mn-O (b) lengths, and tilt angle (c) on the Y-content. In figure 4(b) the *l* and *s* denotes larger and smaller.

Figure 5. Raman spectra of $Eu_{1-x}Y_xMnO_3$, for $x$ = 0, 0.1, 0.3, 0.4 and 0.5, recorded at room temperature. The laser plasma line is indicated by (*).

Figure 6. Dependence of the frequency of the symmetric stretching mode (SS) (a), bending mode (B) (b), antisymmetric stretching mode (AS) (c) and tilt mode (T) (d), on the Y-content. The solid lines are guides for the eyes.

Figure 7. Variation with the tilt angle of the frequency of the *T* mode. The solid line was obtained from the best fit of the linear function to the experimental data.

Figure 8. The ratio between the intensities of the bending and asymmetric stretching modes, as a function of the Y-content. Inset: linewidth of the symmetric stretching mode versus Y-content.

Figure 9. The Raman spectra of $EuMnO_3$ (a), $Eu_{0.8}Y_{0.2}MnO_3$ (b) and $Eu_{0.5}Y_{0.5}MnO_3$ (c), recorded at 200K and at 9K. The laser plasma line is indicated by (*).

Figure 10. Raman spectra of $Eu_{0.8}Y_{0.2}MnO_3$, recorded at several fixed temperatures in the 590-640 cm-1 spectral range. The solid lines were obtained from the best fit of Eq. 3 to the spectra.

Figure 11. Temperature dependence of the Raman frequency of the symmetric stretching mode of $Eu_{1-x}Y_xMnO_3$. The insets show the temperature dependence of the linewidths; the solid lines have been obtained from the best fit of Eq. 5 and 6 (see text) to the experimental data for T > 100K.

Figure 12. Isothermal magnetization as a function of the applied magnetic field, recorded at various temperatures.

Figure 13. Frequency deviation from the extrapolated temperature behaviour for T>100K, of the symmetric stretching mode of $Eu_{0.8}Y_{0.2}MnO_3$, versus the spectral weight calculated from the terahertz data, below 140 $cm^{-1}$, for the sample $Eu_{0.75}Y_{0.25}MnO_3$ (Ref. 17). The straight line is a guide for the eyes.

Table Captions :

Table I. $\Delta_d$ parameter (see Eq. 4) for some rare-earth manganites. The values were obtained from Ref. 23.

| Rare-earth ion | $\Delta_d \times 10^4$ |
| --- | --- |
| La | 33.1 |
| Pr | 43.1 |
| Dy | 49.7 |
| Er | 43.0 |

Table I

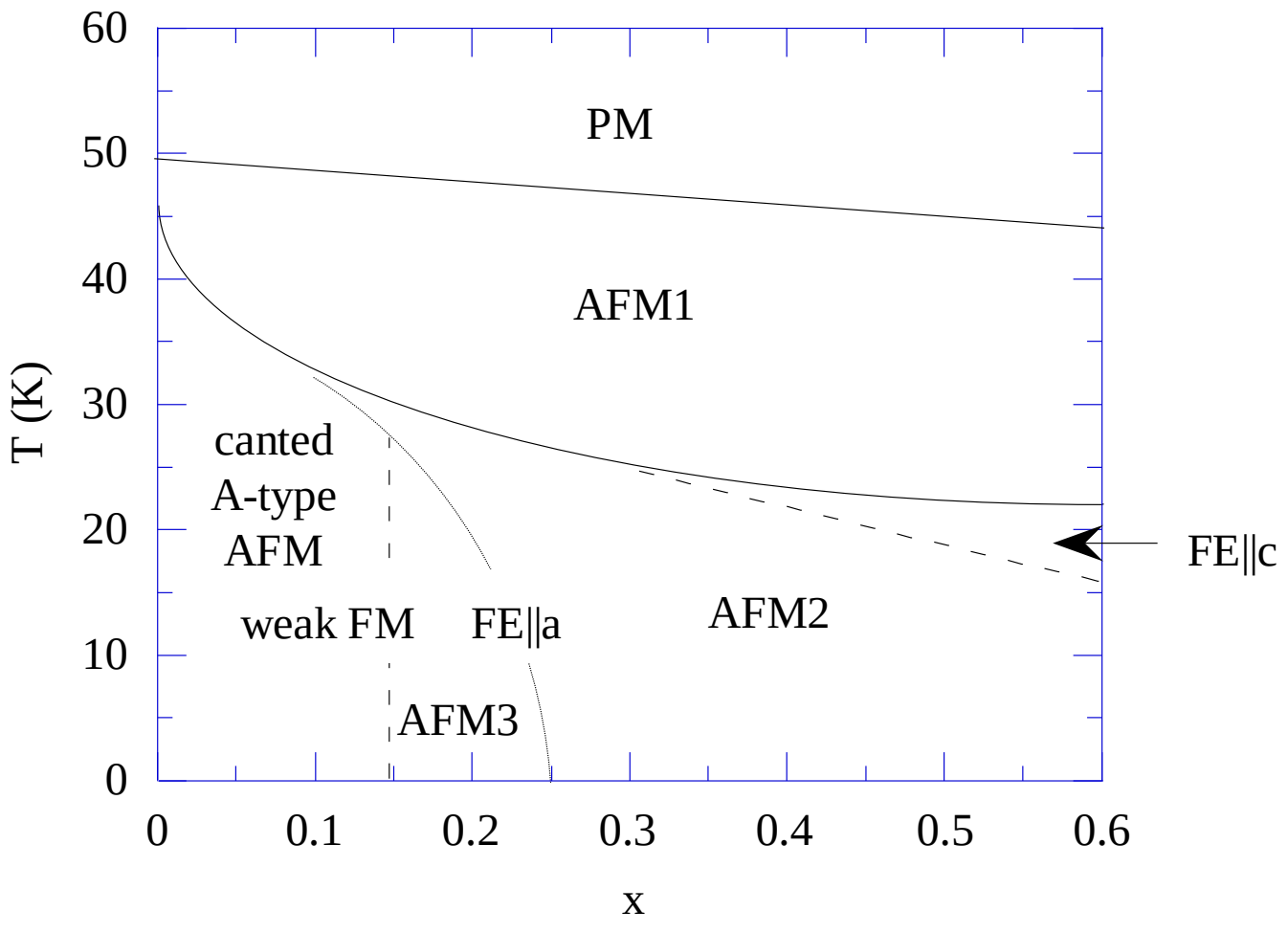

Figure 1

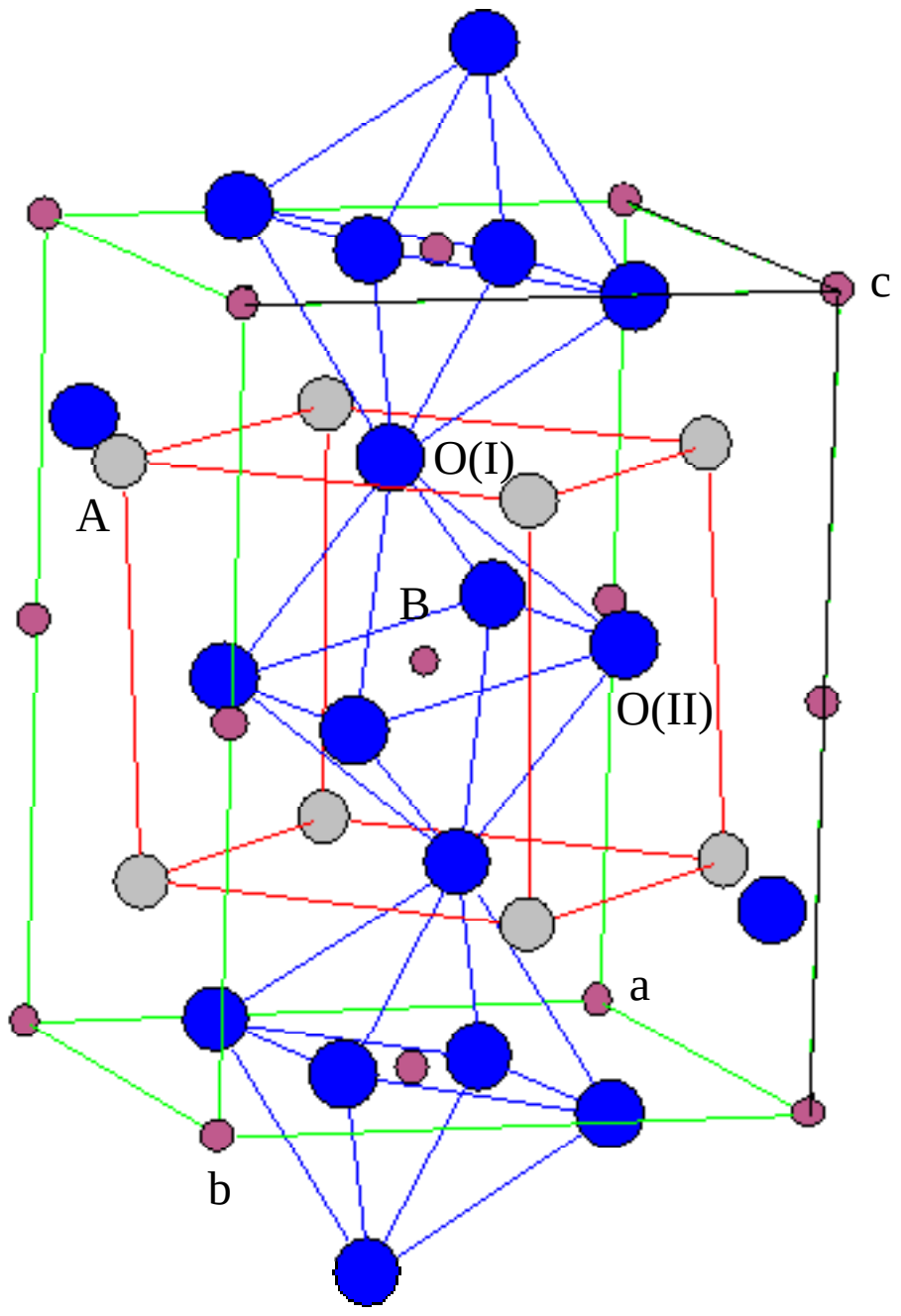

Figure 2

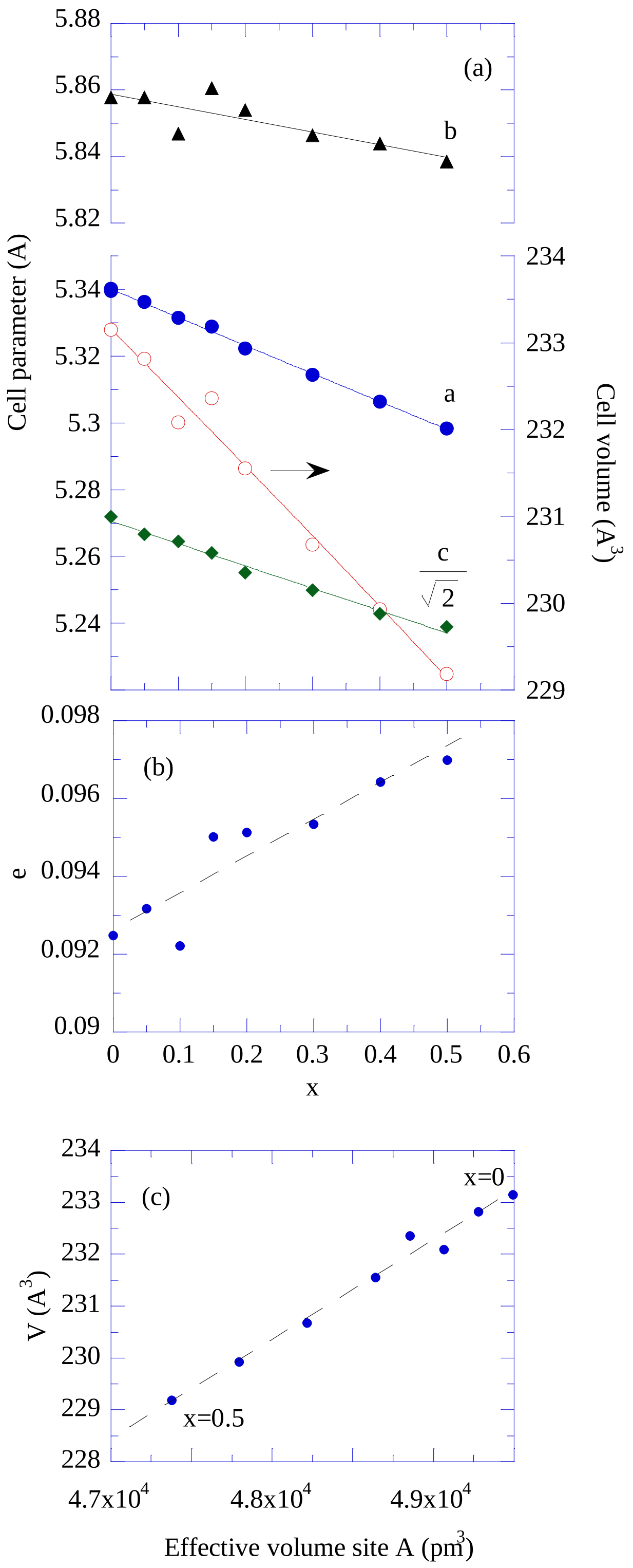

Figure 3

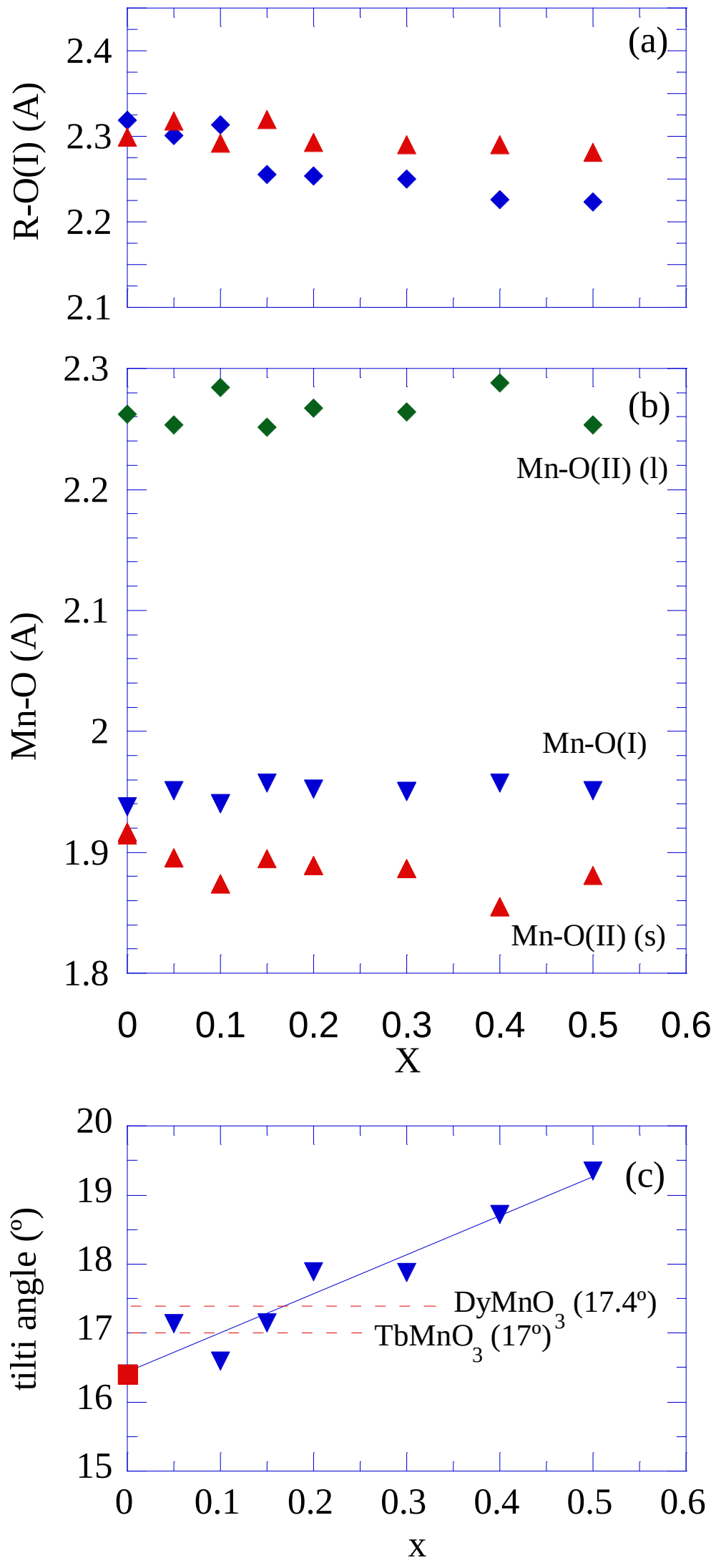

Figure 4

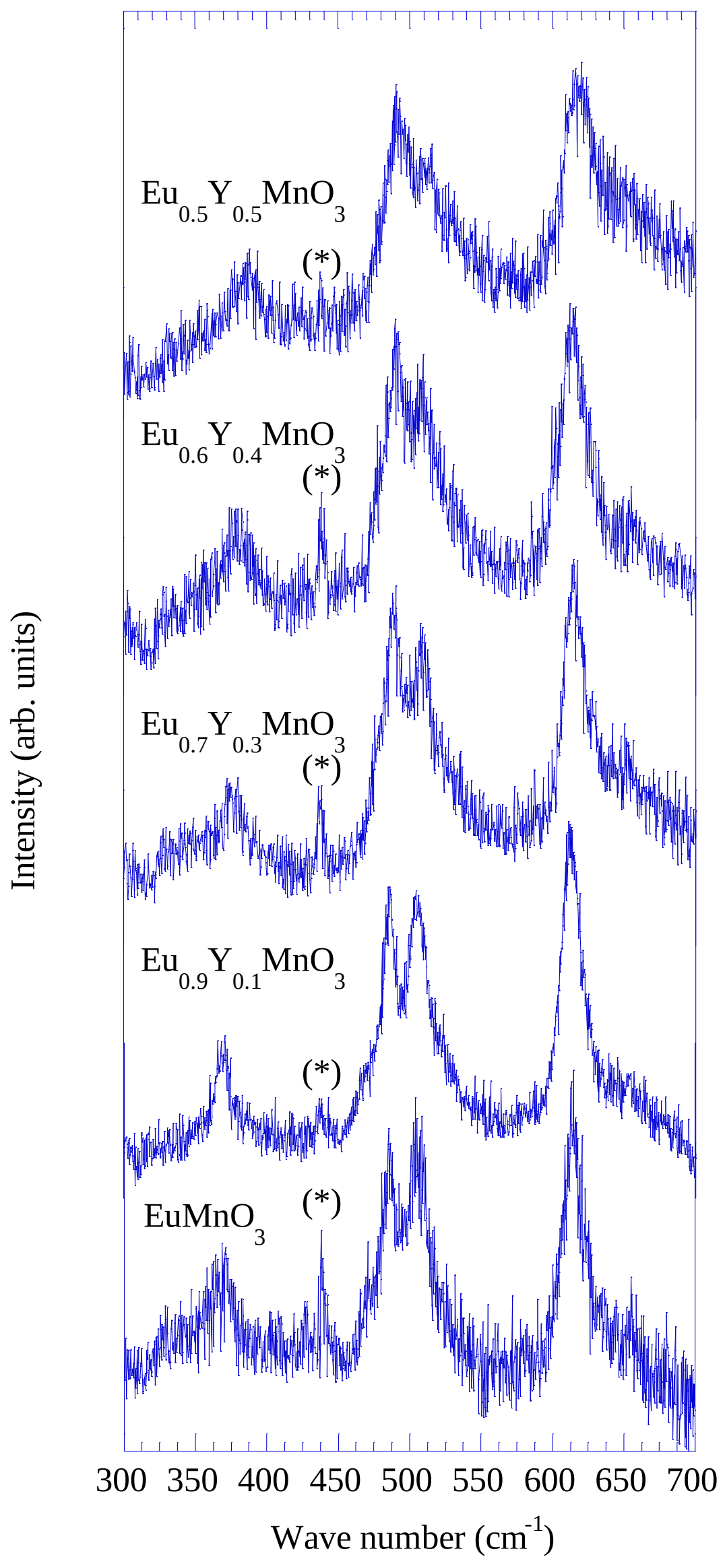
Eu0.5Y0.5MnO3
(*)
Eu0.6Y0.4MnO3
(*)
Eu0.7Y0.3MnO3
(*)
Eu0.9Y0.1MnO3
(*)
EuMnO3
(*)
Intensity (arb. units)
300 350 400 450 500 550 600 650 700
Wave number (cm-1)

Figure 5

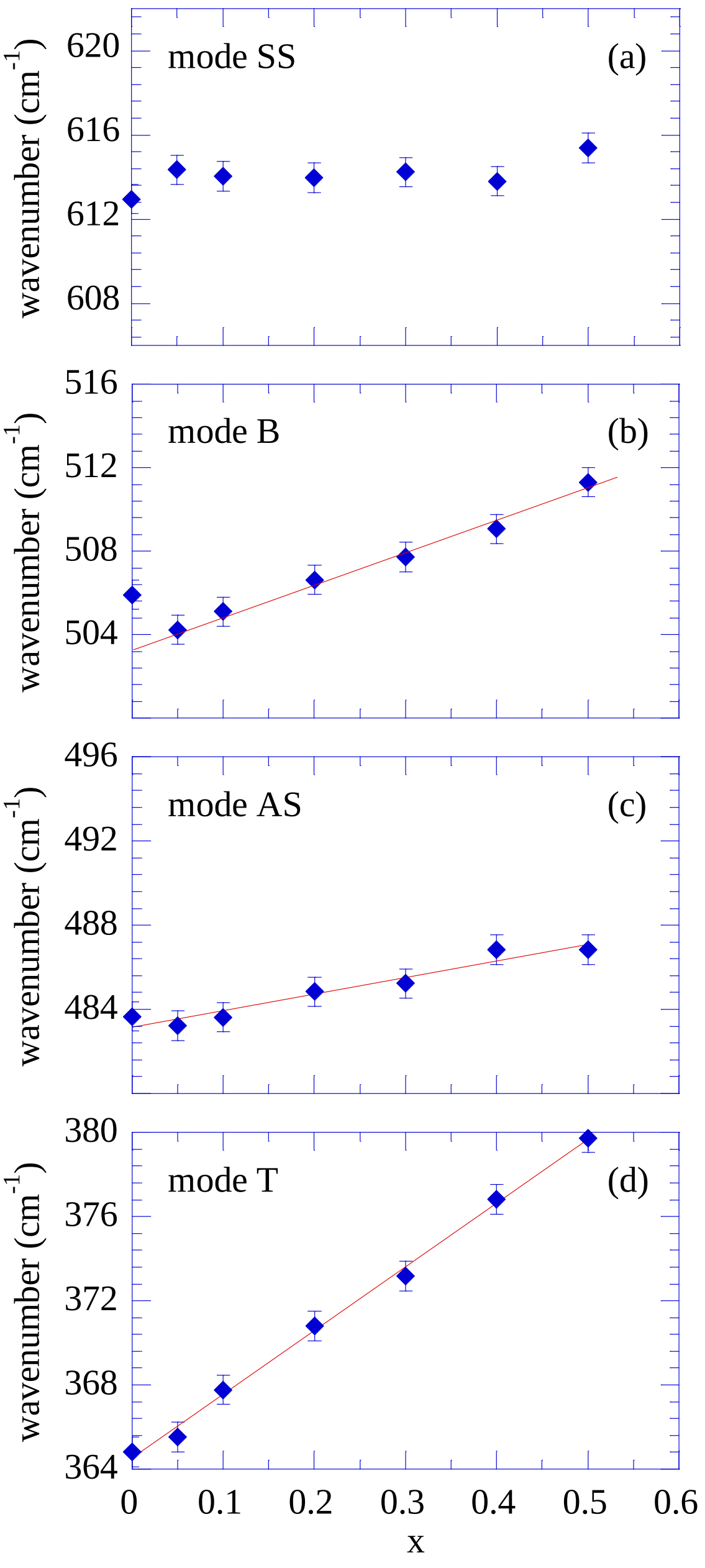

Figure 6

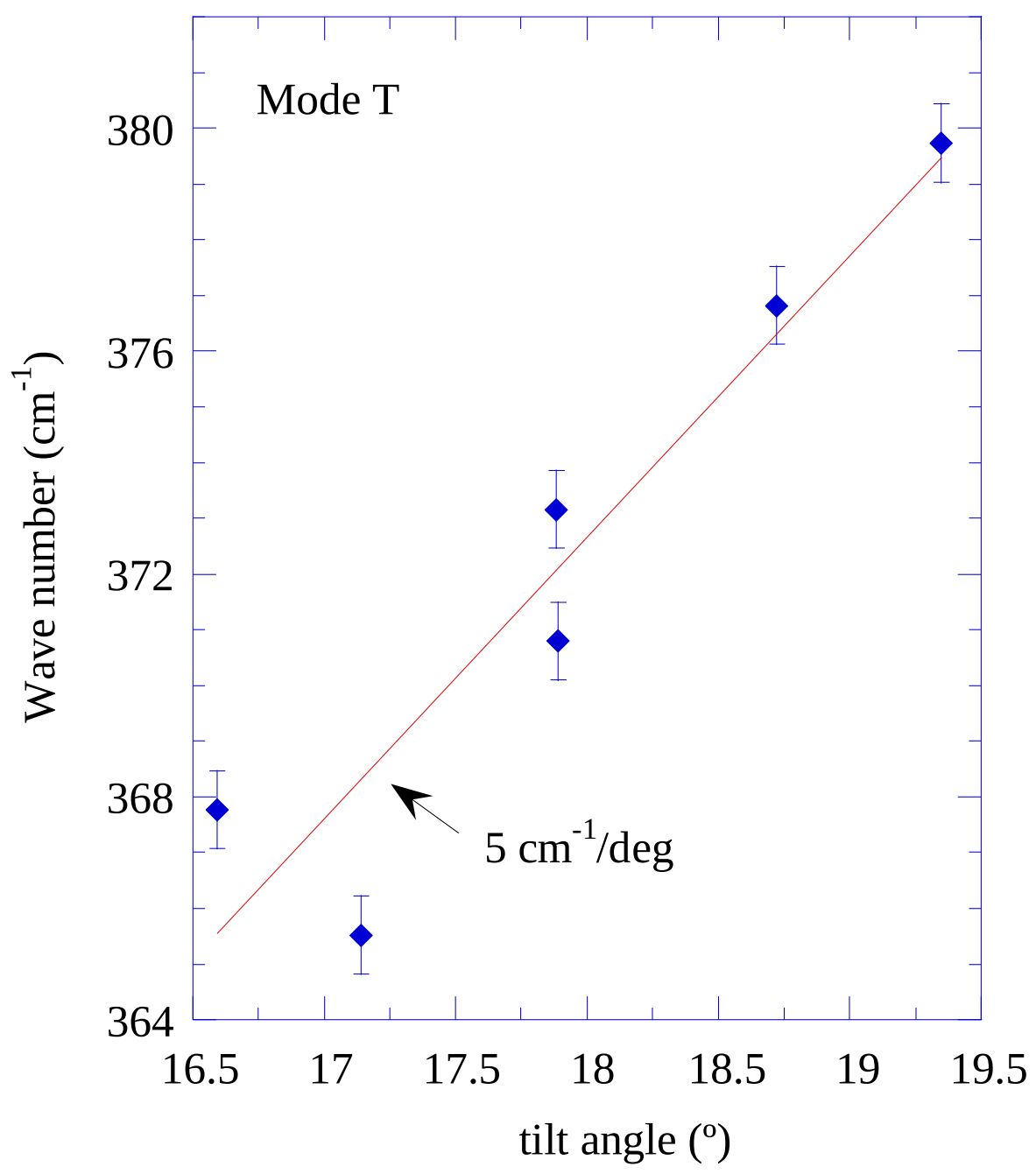

Figure 7

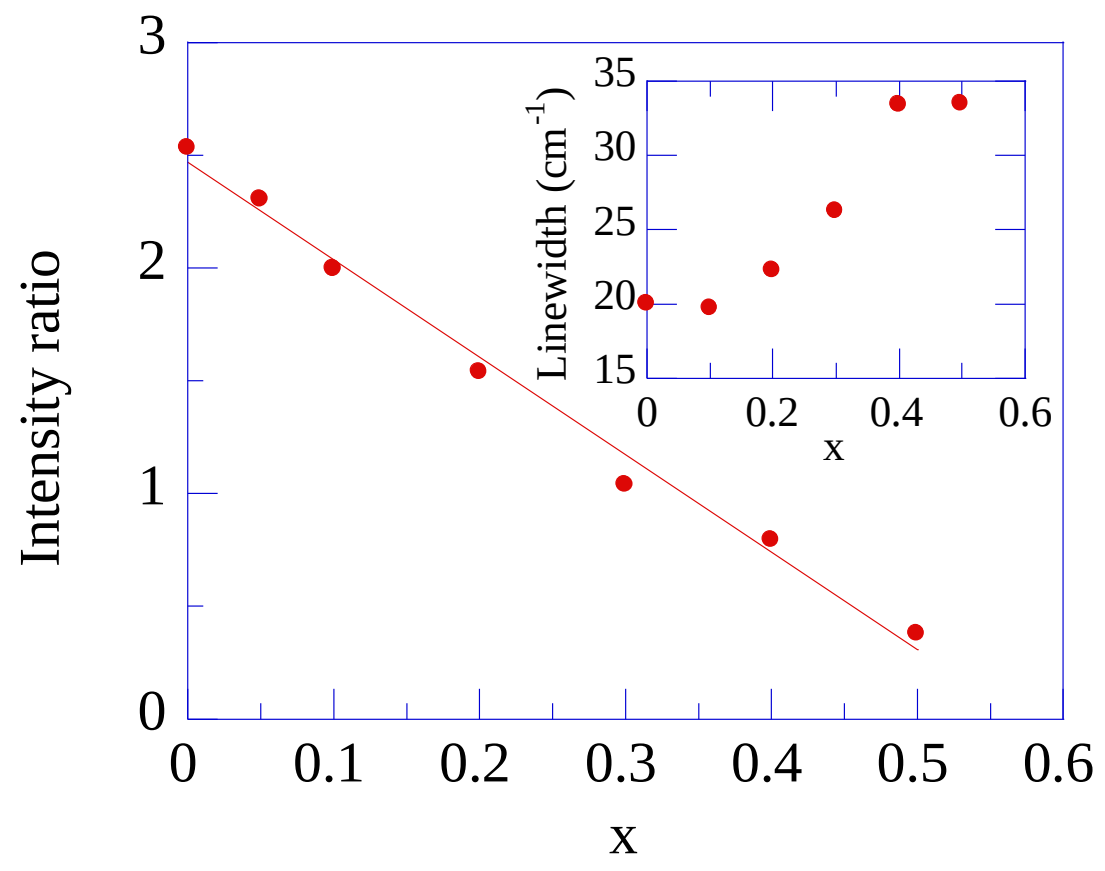

Figure 8

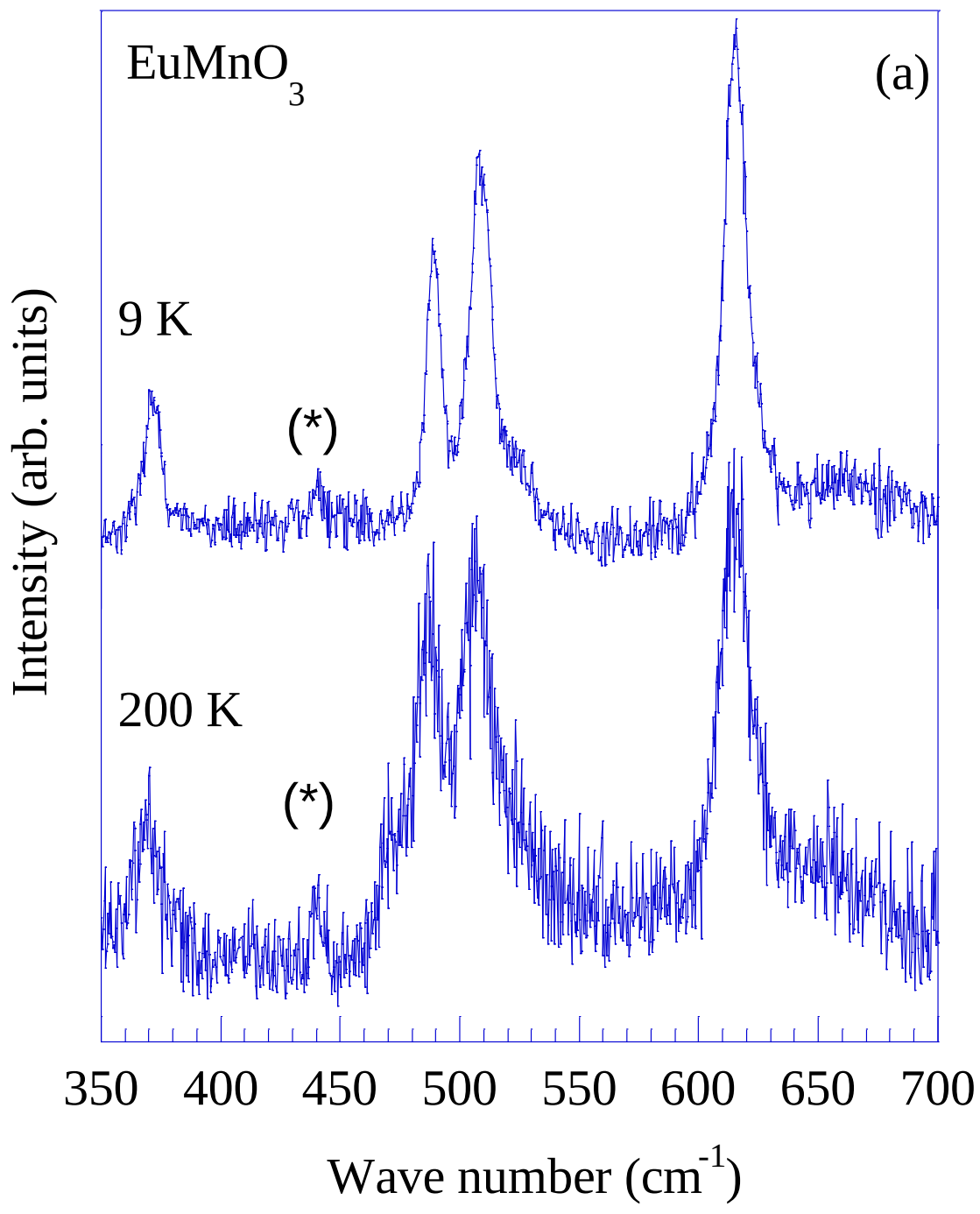

Figure 9(a)

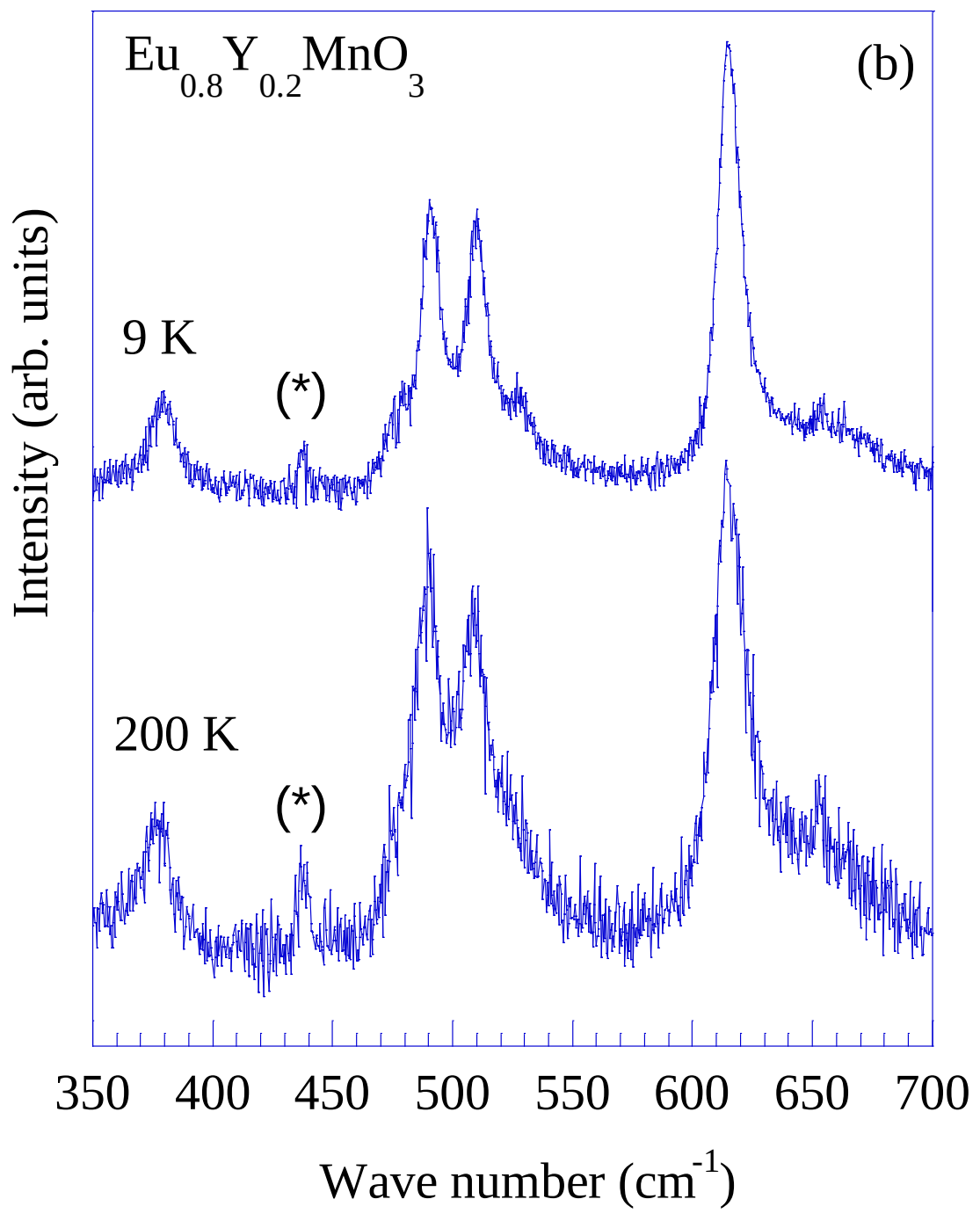

Figure 9(b)

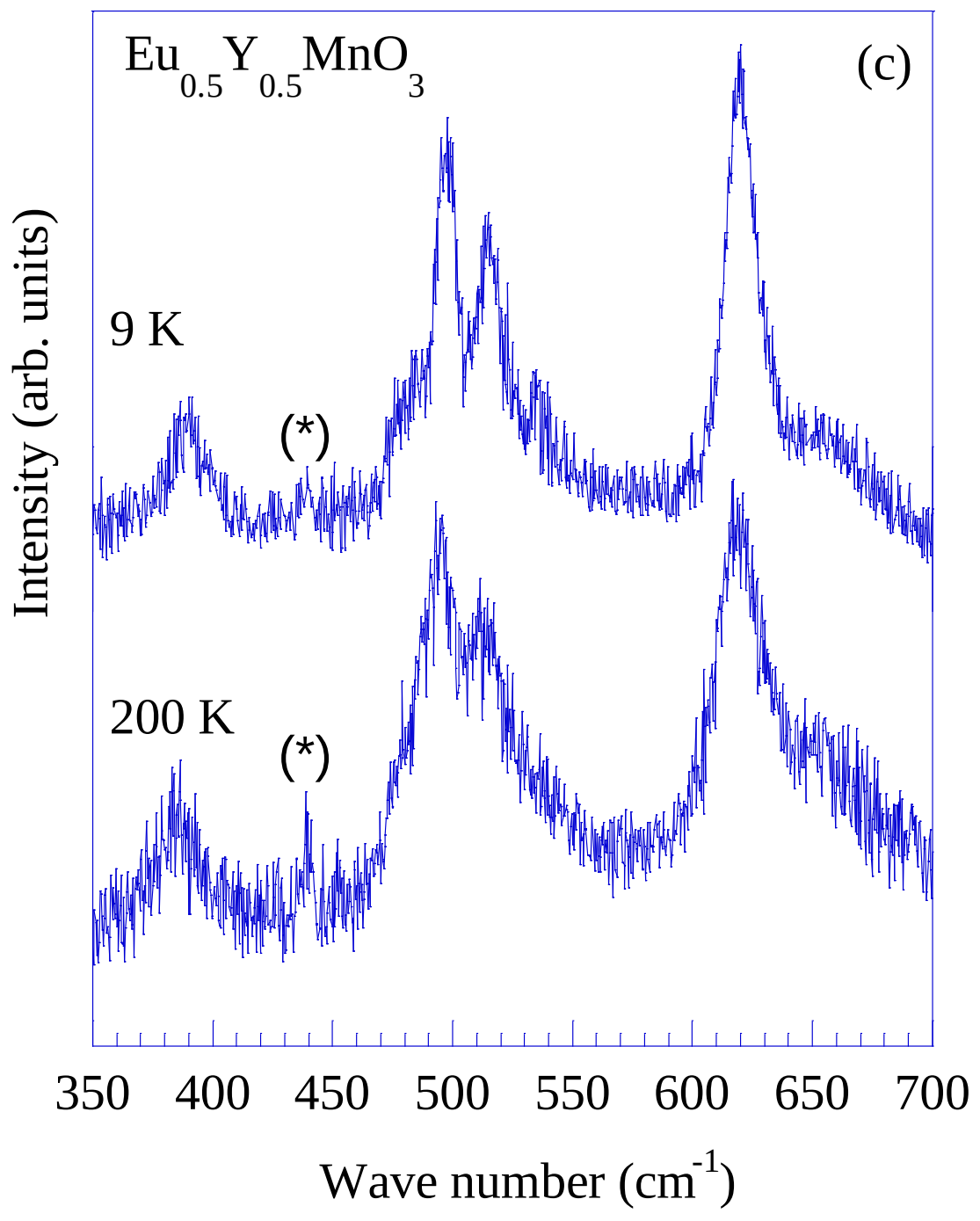
Eu0.5Y0.5MnO3
(c)
9 K
(*)
200 K
(*)
Intensity (arb. units)
350 400 450 500 550 600 650 700
Wave number (cm-1)

Figure 9(c)

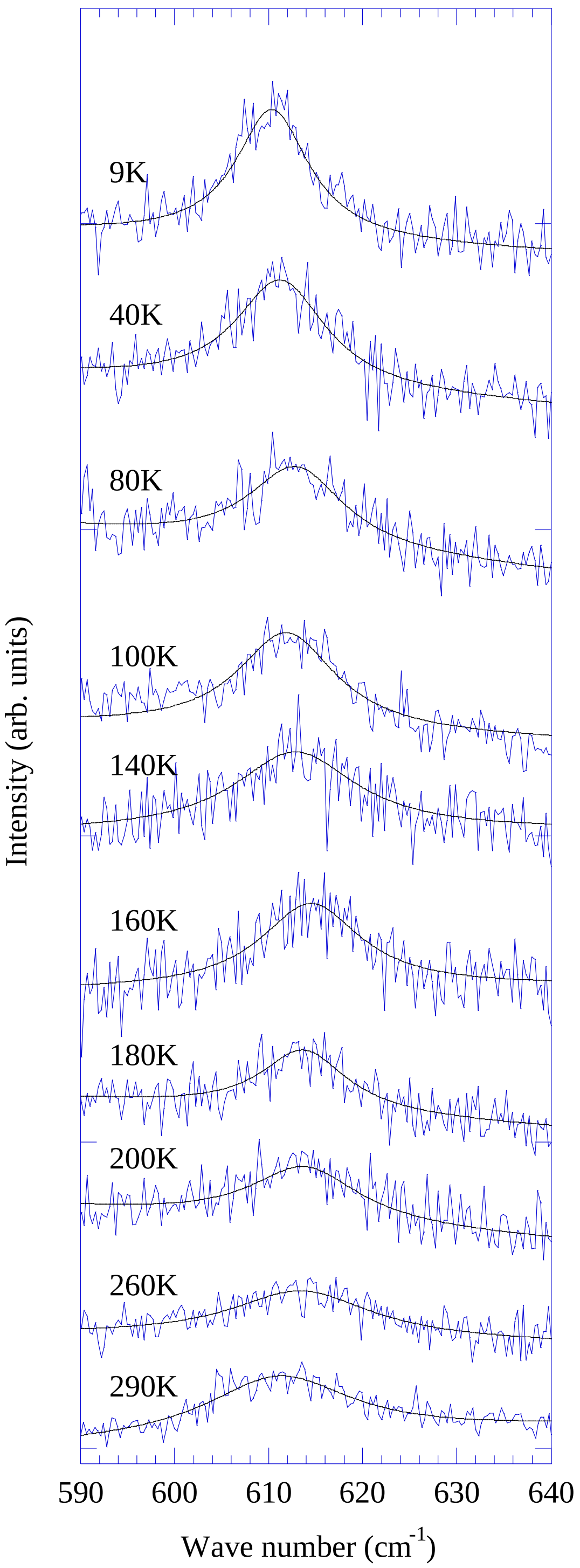

Figure 10

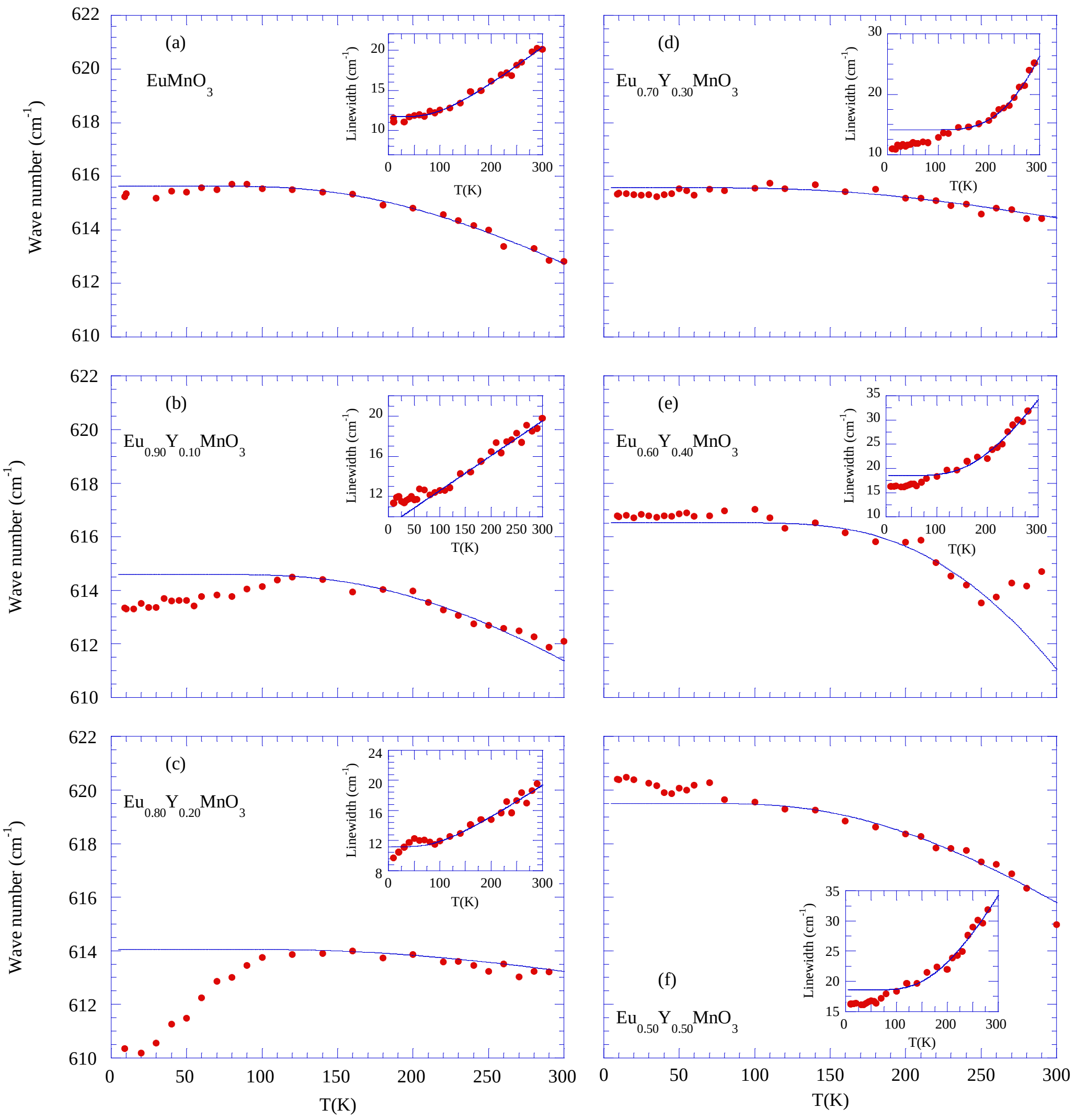

Figure 11

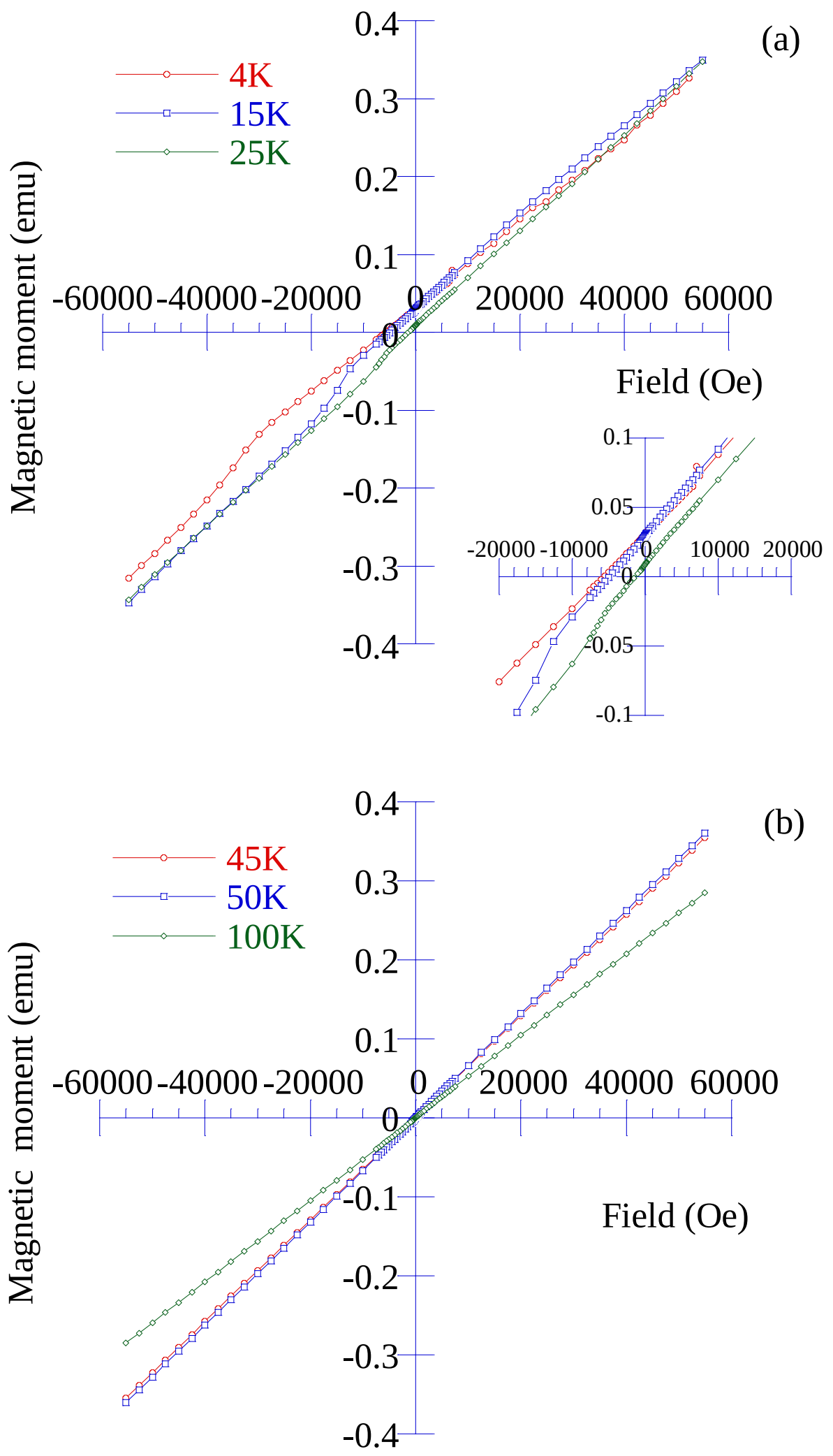

Figure 12

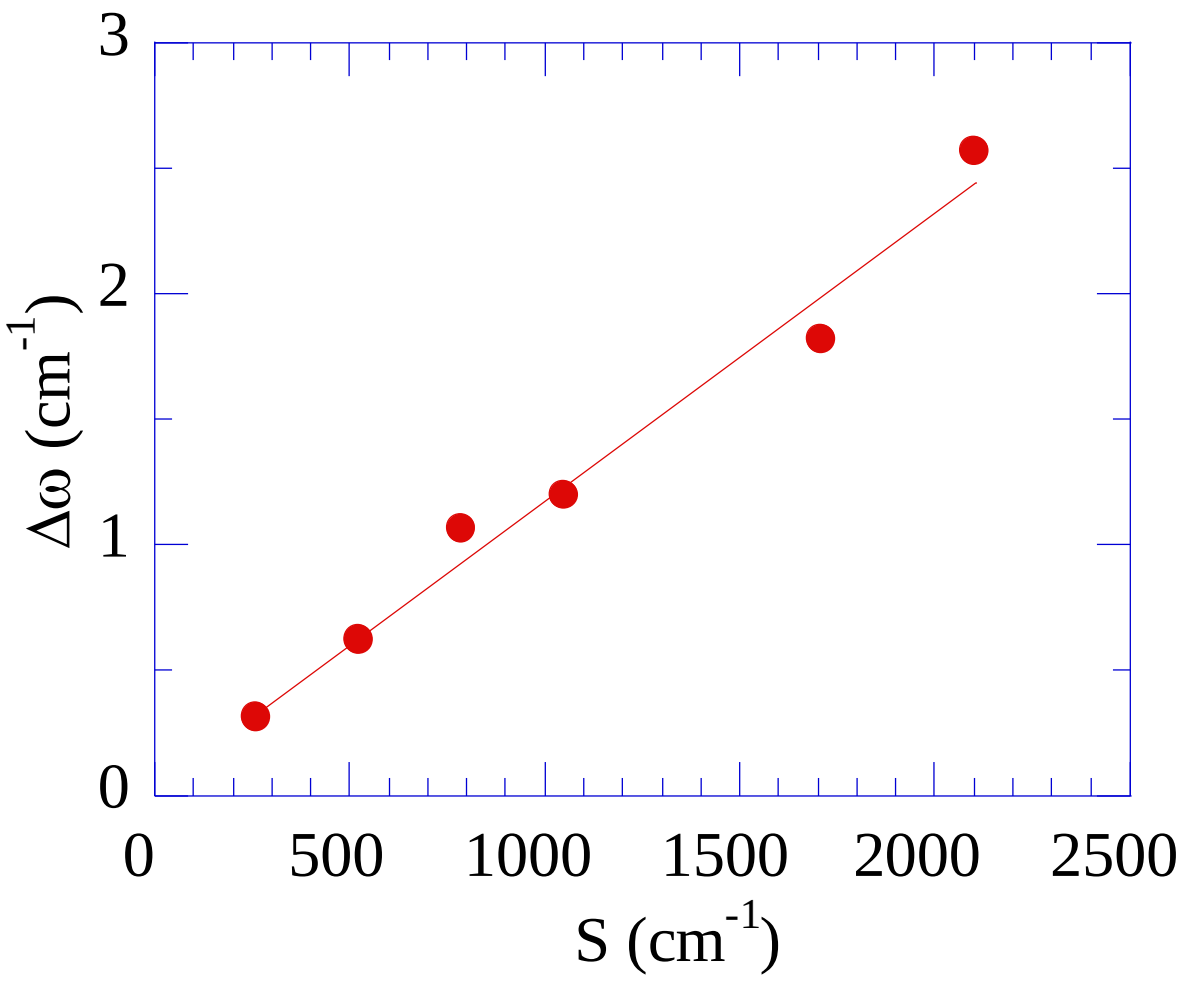

Figure 13